\documentclass[aps,twocolumn,amssymb,showpacs]{revtex4}

\usepackage{graphicx}

\newcommand{\be}{\begin{equation}}
\newcommand{\ee}{\end{equation}}
\newcommand{\bea}{\begin{eqnarray}}
\newcommand{\eea}{\end{eqnarray}}

\def\p1{\pi_1}

\begin{document}

\title{Cutoff AdS/CFT duality and the quest for braneworld
black holes}
\author{Paul R. Anderson$^{\rm a}$}
\altaffiliation{\tt anderson@wfu.edu} 
\author{R. Balbinot$^{\rm b}$}
\altaffiliation{\tt balbinot@bo.infn.it} 
\author{A. Fabbri$^{\rm b,\ c}$}
\altaffiliation{\tt fabbria@bo.infn.it}
\affiliation{${}^{a)}$Department of Physics, Wake Forest University,
Winston-Salem, NC 27109, USA,}
\affiliation{$^{b)}$Dipartimento di Fisica dell'Universit\`a di Bologna 
and INFN sezione di Bologna, Via Irnerio 46, 40126 Bologna, Italy,}
\affiliation{$^{c)}$ Departamento de F\'{\i}sica Te\'orica and IFIC, 
Facultad de 
F\'{\i}sica, Universidad de Valencia, Burjassot-46100, Valencia, Spain.}

\begin{abstract}
We present significant evidence in favour of the 
holographic
conjecture that ``4D
black holes localized on the brane found by solving the classical
bulk equations in $AdS_5$ are quantum corrected black holes and
not classical ones''. The crucial test is the calculation of the
quantum correction to the Newtonian potential based on a numerical
computation of $\langle T^a_{\ b}\rangle $ in
Schwarzschild
spacetime for matter fields in the zero temperature Boulware vacuum
state. For the case of
the conformally invariant scalar field
the leading order term is found to be
$M/45\pi r^3$.
This result is equivalent to the result
which was
previously obtained in the weak-field
approximation using Feynman diagrams and
which has been shown to
be equivalent, via the AdS/CFT duality, to the analogous
calculation in Randall-Sundrum braneworlds.
This asymptotic behavior was not captured in the analytical approximations for
$\langle T^a_{\ b}\rangle $ proposed in the literature.
The 4D backreaction equations are then used to make a prediction about the 
existence and the possible spacetime structure of macroscopic static 
braneworld black 
holes. 

\end{abstract}


\maketitle

The possibility of relating seemingly different theories via
duality relations is a powerful tool which
allows known results in one theory to be used to predict the outcome of
difficult computations in the other. In recent years growing
attention has been devoted to the so called AdS/CFT correspondence
\cite{maldacena}, which
predicts a one-to-one correspondence between a quantum gravity
theory in anti-de Sitter (AdS) space and a conformal field theory
(CFT) living in its boundary at infinity. A variant of this
duality was proposed in \cite{molti, duff} to allow
for
the possibility
that $AdS$ space is cut-off at some finite distance $L$ (the $AdS$
length).
This happens in the RS2 braneworld model \cite{rs2}
where our universe is seen as a hypersurface, the boundary-brane,
which is
immersed in $AdS_5$ space, the bulk. The presence of the brane has
two
primary consequences: i) the dual CFT is cut off at the scale
$1/L$; ii) the zero-mode of 5D gravity gets trapped on the brane
reproducing 4D gravity, which is then added to the dual CFT. The
holographic interpretation of the Randall-Sundrum braneworlds
states that the dual of the classical bulk theory is a CFT, more
specifically, ${\cal N}=4$ SU(N) super Yang-Mills theory in the
large N planar limit, coupled to 4D gravity. In the study of quantum 
properties of matter-gravity systems, a widely used approach 
(semiclassical gravity) consists in treating gravity 
classically
using general relativity and coupling it to
quantum matter fields
via the expectation value of the stress-energy tensor operator for the fields.
It appears then very natural to compare 4D semiclassical
results
with 5D braneworld
results and viceversa.

It was conjectured in \cite{efk} that for large mass black
holes 
{\it 4D black holes localized on the brane found by
solving the classical bulk equations in $AdS_5$ are quantum
corrected black holes and not classical ones}. 
If correct, this 
holographic
conjecture
opens a new perspective for the study of quantum effects in black
hole spacetimes (for instance the information loss problem) using 5D
classical bulk physics.

The 
holographic
conjecture explains 
the results of \cite{bgm}, where it was shown that it is
not possible to match a collapsing sphere of dust on the brane
with a static exterior. According to the conjecture, the
deviations from staticity can be explained in terms of the Hawking
radiation, which introduces time dependence into the system.
It is important to mention that the conjecture is in excellent agreement
with the exact solutions found for black holes localised in a 2+1 brane in 
$AdS_4$ \cite{ehm}.
However, in $AdS_5$, where despite much effort 
\cite{tanti} black hole solutions 
have not been found, no actual proof of 
the 
holographic
conjecture 
has been provided so far.  

We give here a new and important check of this conjecture by using the 
semiclassical backreaction equations to compute the quantum corrected 
4D Newtonian potential and showing that to leading order it is equivalent 
to that obtained classically in the 
$AdS_5$ bulk in the weak field limit.  We then use the
conjecture along with the well known properties of the stress-energy tensor
for quantized free fields in the zero temperature Boulware state 
to make a
prediction regarding the existence of static black hole solutions to the bulk
equations. 

In the weak field limit
it has been shown in \cite{duff} that the
Randall-Sundrum result for the gravitational potential \cite{gata} \be 
\Phi =
\frac{M}{r}(1+\frac{2}{3}\frac{L^2}{r^2}) \ee is equivalent to the
4D computation based on the one-loop quantum corrections to the
graviton propagator \cite{dudo}
due to conformally invariant fields,
which gives \be \label{corrgrav} \Phi=
\frac{M}{r}(1+\frac{\alpha}{r^2})\ . \ee The value of $\alpha$
depends on the specific
numbers and types of fields considered \be
\alpha=\frac{1}{45\pi}(12N_1 + 3N_{1/2} + N_0)\ .\ee
Here subscripts correspond to the spin of the field.
The matching
between the two expressions
for the potential
requires
a specification of the number of
degrees of freedom for each matter field species of the particular
dual CFT theory, i.e.,  $N_1=N^2,\ N_{1/2}=4N^2,\ N_0=6N^2$,
along with the relation $N^2=\pi L^2$
which is
derived from
the
AdS/CFT
correspondence
combined with the
Randall-Sundrum formula involving five dimensional and four
dimensional Newton's constants (see \cite{duff} for more details).

In this paper, using semiclassical gravity, we derive the four dimensional 
gravitational potential 
by first computing the stress-energy
tensor for a conformally invariant scalar field in the Boulware state 
\cite{boul} in
Schwarzschild spacetime.  The leading order behavior of this stress-energy tensor
is obtained in the region far from the event horizon.  Then the linearized semiclassical
backreaction equations are integrated 
to obtain the quantum corrected Newtonian potential. The end result is 
then compared to Eq. (\ref{corrgrav}) to check the conjecture. 

The stress-energy tensor for
the
conformally coupled
massless scalar field
in the Boulware state
 has previously been numerically computed in
\cite{jmo} and in  \cite{bo} analytic approximations  have been
computed. Moreover, in \cite{ahs} an analytic approximation has
been
derived
that can be used to obtain an approximation for
the stress-energy tensor for arbitrarily coupled massless scalar
fields in the Boulware state.

The analytic approximations predict that at large values of the
radial coordinate $r$ the nonzero components of the stress-energy
tensor have leading order behaviors that are proportional to
$M^2/r^6$, with $M$ the mass of the black hole. It is clear that
such
a
term cannot reproduce the correction in Eq. (\ref{corrgrav})
(in fact, it generates a quantum correction to $\Phi$ of the order
$M^2/r^4$).
This presents a 
serious
challenge for the holographic conjecture. 
One way to
resolve this
issue
is 
to compute the stress-energy tensor numerically.
As mentioned above this has been done in Ref.~\cite{jmo}.
However it is not possible to deduce the
large $r$ behavior from the plots of the numerical results in
that paper.

We have numerically computed the stress-energy tensor for massless
scalar fields with arbitrary coupling to the scalar curvature in the
Boulware state in Schwarzschild spacetime.  The method used
is the same as that given in Ref.~\cite{ahs}, which in turn
is an adaptation and generalization of the method originally used
in \cite{ch1,ch2,howard} for the conformal scalar field in
Schwarzschild spacetime. Renormalization is accomplished through the use
of point splitting.  In principle one can subtract the point
splitting counterterms computed in \cite{christensen} from the
unrenormalized stress-energy tensor and then take the limit as the
points come together.  In practice it is easier to add and
subtract terms using the WKB approximation for the radial modes.
As shown in Ref.~\cite{ahs} it is possible
to use the high frequency limit of the WKB approximation to write the
stress-energy tensor in terms of two finite tensors $\langle T_{ab} \rangle_{\rm numerical}$ and
$\langle T_{ab} \rangle_{\rm analytic}$ that are separately
conserved.  The result is
\begin{eqnarray}
\langle T_{ab} \rangle_{\rm ren} &=& \langle T_{ab} \rangle_{\rm
numerical} + \langle T_{ab} \rangle_{\rm analytic} \\ \langle
T_{ab} \rangle_{\rm numerical} &=& \langle T_{ab} \rangle_{\rm
unren} - \langle T_{ab} \rangle_{\rm WKBdiv} \nonumber \\ \langle
T_{ab} \rangle_{\rm analytic} &=& \langle T_{ab} \rangle_{\rm
WKBdiv} - \langle T_{ab} \rangle_{\rm ps}  \;. \nonumber
\end{eqnarray}
The second term can be computed
analytically in any static spherically symmetric spacetime and for massless
fields is the analytic approximation derived in~\cite{ahs} which is mentioned
above.

To actually compute the stress-energy tensor numerically it is
useful to add and subtract the full WKB approximation 
with the result that
\begin{eqnarray}
\langle T_{ab} \rangle_{\rm numerical} &=& \langle T_{ab}
\rangle_{\rm modes} + \langle T_{ab} \rangle_{\rm WKFfin} \\
\langle T_{ab} \rangle_{\rm modes} &=& \langle T_{ab} \rangle_{\rm
unren} - \langle T_{ab} \rangle_{\rm WKB} \nonumber \\ \langle
T_{ab} \rangle_{\rm WKBfin} &=& \langle T_{ab} \rangle_{\rm WKB} -
\langle T_{ab} \rangle_{\rm WKBdiv}  \;. \nonumber
\end{eqnarray}
It turns out that $\langle T_{ab} \rangle_{\rm modes}$ and
$\langle T_{ab} \rangle_{\rm WKFfin} $ are not separately
conserved.  However, for a zero temperature massless scalar field it is possible
to compute the latter analytically (except for a few integrals
that must be computed numerically) for an arbitrary static
spherically symmetric spacetime.  
However, the mode sums 
converge much more rapidly than they do if one computed the quantity
$\langle T_{ab} \rangle_{\rm numerical} $ directly.  The higher
the order of the WKB approximation the faster the mode sums
converge.  In the calculations below an eigth order WKB expansion
was used in $\langle T_{ab} \rangle_{\rm WKB}$.

Because the scalar curvature is zero in Schwarzschild spacetime, it
is possible to write the stress-energy tensor for arbitrary
coupling in the general form~\cite{ahs}
\begin{equation}
\langle T_{ab} \rangle = C_{ab} + (\xi - \frac{1}{6}) D_{ab}
\end{equation}
with $\xi$ the coupling to the scalar curvature. The numerical
results for the components of the tensors $C_{ab}$ and $D_{ab}$
are shown in Figures 1 and 2.
\begin{figure}
\includegraphics[angle=90,width=3.4in,clip]{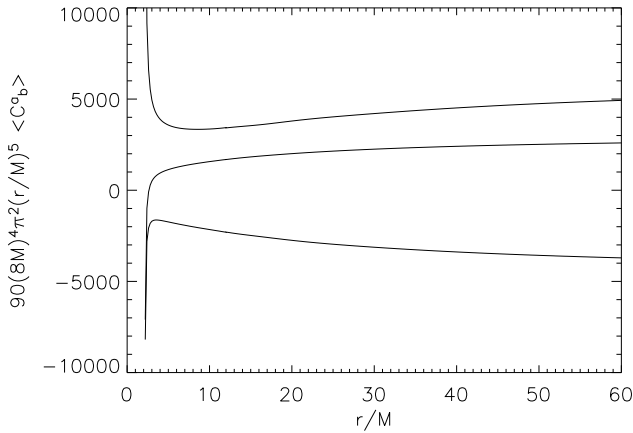}
\caption{The curves from top to bottom at the right of the plot correspond
to ${C^t}_t$, ${C^r}_r$, and ${C^\theta}_\theta$ respectively. } \label{fig1}
\end{figure}
\begin{figure}
\includegraphics[angle=90,width=3.4in,clip]{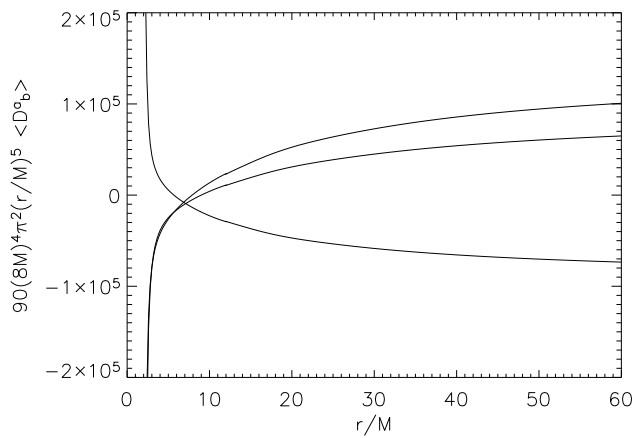}
\caption{The curves from top to bottom at the right of the plot correspond
to ${D^\theta}_\theta$, ${D^t}_t$, and ${D^r}_r$ respectively.} \label{fig2}
\end{figure}
In the figures each component of the stress-energy tensor is
multiplied by a factor of $(r/M)^5$.  It is clear from the plots that
the leading order behavior of the stress-energy tensor at large
$r$ is proportional to $M/r^5$ and not $M^2/r^6$ as predicted by
the analytical approximations.  We find
that to leading order the stress-energy tensor is~\cite{footnote}
\begin{eqnarray} \label{leadord}
\langle T^t_t \rangle &=& \frac{M}{60 \pi^2 r^5}
 + (\xi -\frac{1}{6}) \frac{M}{4 \pi^2 r^5}
\nonumber \\ \langle T^r_r \rangle &=& \frac{M}{120 \pi^2 r^5}
 - (\xi - \frac{1}{6})\frac{M}{4 \pi^2 r^5}
\nonumber \\ \langle T^\theta_\theta \rangle &=& -\frac{M}{80
\pi^2 r^5} + (\xi - \frac{1}{6}) \frac{3M}{8 \pi^2 r^5} \;.
\end{eqnarray}
This tensor
is conserved to
leading order in $1/r$ and for the conformal case, $\xi=1/6$, is
correctly traceless (the trace anomaly being of the order
$M^2/r^6$).

The next step is to determine the quantum corrections to the
Schwarzschild metric and to the Newtonian potential. These can be
computed by solving the semiclassical Einstein equations (backreaction equations),
which by writing
the metric as
\be
ds^2=-(1-2M(r)/r)e^{2\phi(r)}dt^2 +
\frac{dr^2}{\left(1-\frac{2M(r)}{r}\right)}+r^2d\Omega^2 \  \ee
take the simple form  \bea \label{backeq1}
\partial_r M &=& -4\pi r^2 \langle T^t_t \rangle  \ ,  \\
\label{backeq2}
\partial_r\phi &=& -4\pi r \frac{ \langle T^t_t\rangle - \langle T^r_r\rangle
 }{(1-2M/r)}\ . \eea
 At linear order, using the results (\ref{leadord}) we
find that
 \bea \label{schwcorr}
ds^2 &=& -\left(1-\frac{2M}{r}(1+\frac{\alpha}{r^2}) \right)dt^2
\nonumber \\ &+& \left( 1- \frac{2M}{r}(1+ \frac{\beta}{ r^2})\right)^{-1}
dr^2 +r^2d\Omega^2 \ ,\eea where \be \label{alphabeta}
\alpha= \frac{1}{45\pi} - (\xi-\frac{1}{6})\frac{1}{6\pi},\ \ \
\beta= \frac{1}{30\pi} + (\xi-\frac{1}{6})\frac{1}{2\pi}\ .\ee The
quantum correction to the Newtonian potential takes the form of Eq.\
(\ref{corrgrav}) with $\alpha$ given in Eq.\ (\ref{alphabeta}). For
the conformal case $\xi=1/6$ we exactly reproduce the effective
potential calculation in Eq.~(\ref{corrgrav}).  For the minimally coupled case, $\xi=0$, Eq. (\ref{alphabeta})
gives $\alpha=1/20$ which reproduces the analogous computation with
Feynman diagrams performed in \cite{hamliu}.
To complete the proof that the classical bulk Newtonian potential is equivalent
to the 4D quantum corrected potential one should also show
the matching of the results for
massless
spin $1/2$ and spin $1$ fields.
However,
the agreement found for the conformal and minimally coupled scalar fields
makes it reasonable
to suppose that the agreement will extend to nonzero spin fields
as well~\cite{ch}.
It would be interesting to improve the existing analytic
approximations for the stress-energy tensor in the
Boulware state in order to reproduce the results
in Eq.~(\ref{leadord}).

The importance of our result is twofold. On one hand, it
provides the first proof that calculations of the effective potential
which make use of Feynman diagrams to
compute corrections to the graviton propagator give the same answer as
that obtained by computing the stress-energy tensor for the quantized fields and
solving the linearized semiclassical backreaction equations.
In addition, and this is the main focus of the present paper, it
provides a strong check of the holographic interpretation for
braneworld black holes, which
makes an
identification between classical solutions of black holes localised on the 
brane and
solutions to the semiclassical backreaction equations in 4D
black hole spacetimes.

The holographic interpretation
 is quite important because in principle the semiclassical
Einstein equations (\ref{backeq1}), (\ref{backeq2})
allow
one
to determine
not only
the spacetime metric in the asymptotic region at
large values of $r$, Eq. (\ref{schwcorr}), but also
the possible spacetime structure at intermediate values of $r$. It is
well known \cite{chrisfull} that the requirements that the stress 
tensor is static and
vanishes asymptotically (which are equivalent to demanding a 
zero
temperature Boulware  state for the matter fields) imply that
$\langle T^a_{\ b}\rangle$ strongly diverges at the classical
horizon of the Schwarzschild spacetime as \be \langle T^a_{\
b}\rangle \sim \frac{(2N_1+\frac{7}{4}N_{1/2}+N_0)}{30\
2^{12}\pi^2 M^4f^2}(1,-1/3,-1/3,-1/3)\  , \ee where $f=1-2M/r$.
Naive insertion of these quantities in the backreaction equations
(\ref{backeq1}), (\ref{backeq2}) gives divergent values for $M(r)$
and $\phi(r)$ in the limit $r\to 2M$. 
One consequence of this result is that in the 
linearized 
approximation we are considering
the classical horizon 
gets destroyed by the quantum corrections.
On the other hand our confirmation of the holographic conjecture in
the weak field limit would seem to imply that some type of nontrivial, static
vacuum solution to the classical bulk equations exists~\cite{emparan}, 
although it is probably
not a black hole.  
Both of these properties can hold only if quantum effects are large
near $r=2M$ for solutions to 
the semiclassical backreaction equations which have the asymptotic behavior~
(\ref{corrgrav}).
For macroscopic black holes one would not expect quantum 
effects to be large near the horizon. The usual interpretation 
of this result is that the Boulware state describes matter fields 
around a static star and not a black hole. 
In fact, the natural thing 
for a black hole is to evaporate via the Hawking 
effect. 
The one possible 
exception would be if the system has charges which allow the presence 
of zero temperature solutions. Indeed, it has been shown that the 
stress energy tensor of massless spin 0 and spin $1/2$ fields in the 
vacuum state is finite in 4D on the event horizon of the (zero-temperature) 
extreme Reissner-Nordstr\"om black hole~\cite{ahl,spin12}.

{\bf Acknowledgements}: We thank S. Fagnocchi, J. Navarro-Salas and in particular R. Emparan and
N. Kaloper for many useful discussions.  This work has been supported in part 
by grant number PHY-0070981 from the National Science Foundation.



\begin{references}
\bibitem{maldacena} O. Aharony, S. S. Gubser, J. Maldacena, H.
Ooguri and Y. Oz, {\it Phys. Rept.} {\bf 323}, 183 (2000)
\bibitem{molti}
J. Maldacena, unpublished; E. Witten, unpublished; H. Verlinde,
{\it Nucl. Phys.} B{\bf 580}, 264 (2000); S. S. Gubser, {\it Phys.
Rev.} D{\bf 63}, 084017 (2001); S. B. Giddings, E. Katz and L.
Randall, {\it J. of High Energy Phys.} {\bf 03}, 023 (2000); S. B.
Giddings and E. Katz, {\it J. Math. Phys.} {\bf 42}, 3082 (2001);
N. Arkani-Hamed, M. Porrati and L. Randall, {\it J. of High Energy
Phys.} {\bf 08}, 017 (2001)
\bibitem{duff} M. J. Duff and J. T. Liu, Phys. Rev. Lett. {\bf 85}, 2052 (2000)
\bibitem{rs2} L. Randall and R. Sundrum, {\it Phys. Rev. Lett.}
{\bf 83}, 4690 (1999)
\bibitem{efk} T. Tanaka, {\it Prog. Theor. Phys. Suppl} {\bf 148},
307 (2003); R. 
Emparan, A. Fabbri and N. Kaloper, {\it J. of High
Energy Phys.} {\bf 08}, 043 (2002); R. Emparan, J. Garcia-Bellido and 
N. Kaloper, {\it J. of High Energy Phys.} {\bf 01}, 079 (2003) 
\bibitem{bgm} M. Bruni, C. Germani and R. Maartens, {\it Phys.
Rev. Lett.} {\bf 87}, 231302  (2001)
\bibitem{ehm} R. Emparan, G. T. Horowitz and R. C. Myers, {\it J. of
High Energy Phys.} {\bf 01}, 007 (2000); {\it J. of High Energy Phys.} 
{\bf 01}, 021 (2000)
\bibitem{tanti} A. Chamblin, S. W. Hawking and H. S. Reall, {\it
Phys. Rev.} D{\bf 61}, 065007 (2000); N. Dadhich, R. Maartens, P. 
Papadopoulos and V. Rezania, {\it Phys. Lett.} {\bf B487}, 1 (2000);
N. Deruelle, {\it Stars on
branes: the view from the brane}, gr-qc/0111065; P. Kanti and 
K. Tamvakis, {\it Phys. Rev.} {\bf D65}, 084010 (2002); R. Casadio, 
A. Fabbri and L. Mazzacurati, {\it Phys. Rev.} D{\bf 65}, 084040
(2002); C. Charmousis and R. Gregory, {\it Class. Quant. Grav.} {\bf 21}, 
527 (2004);
H. Kudoh, T. Tanaka, and T. Nakamura, {\it Phys. Rev.} D{\bf 68}, 024035 (2003);
H. Kudoh,  {\it Phys. Rev.} D{\bf 69} , 104019 (2004); G. Kofinas, E. Papantonopoulos, and
V. Zamarias, {\it Phys. Rev.} D{\bf 66}, 104028 (2002).
\bibitem{gata}
J. Garriga and T. Tanaka, {\it Phys. Rev. Lett.} {\bf 84}, 2778
(2000)
\bibitem{dudo}
M. J. Duff, {\it Phys. Rev.} {\bf D9}, 1837 (1974); J. F. Donoghue,
{\it Phys. Rev. Lett.} {\bf 72}, 2996 (1994)
\bibitem{boul}
D. G. Boulware, {\it Phys. Rev.} {\bf D11}, 1404 (1975)
\bibitem{jmo} B. P. Jensen, J. G. Mc Laughlin, and A. C. Ottewill,
Phys. Rev. D{\bf 45}, 3002 (1992).
\bibitem{bo} M. R. Brown and A. C. Ottewill, Phys. Rev. D{\bf 31}, 
2514 (1985); V. P. Frolov and A. I. Zelnikov, {\it Phys. Rev.} {\bf D35},
3031 (1987)
\bibitem{ahs} P. R. Anderson, W. A. Hiscock, and D. A. Samuel, Phys. Rev. D
{\bf 51}, 4337 (1995).

\bibitem{ch1} P. Candelas and K. W. Howard, Phys. Rev. D {\bf 29}, 1618 (1984).

\bibitem{ch2} K.W. Howard and P. Candelas, Phys. Rev. Lett. {\bf 53}, 403 (1984).

\bibitem{howard} K.W.Howard, Phys. Rev. D {\bf 30}, 2532 (1984).

\bibitem{christensen} S. M. Christensen, Phys. Rev. D{\bf 14}, \rm 2490 (1976)

\bibitem{footnote} The analytic values displayed in Eq. (\ref{leadord}) have actually been obtained through
a combination of computing analytically certain sums~\cite{ahs} that contribute to the
quantity $\langle T_{ab} \rangle_{WKBfin}$ and using conservation to 
obtain the
rest.  The details of this procedure will be described elsewhere.  When a fitting 
routine from the algebraic manipulation program Mathematica 
is used for the same sums it is found that there is agreement between the analytic and numerical
values for the coefficients of the $M/r^5$ terms in $\langle T_{ab} \rangle$ to within 
approximately two significant digits.
\bibitem{hamliu} H. W. Hamber and S. Liu, {\it Phys. Lett.} B{\bf
357}, 51 (1995).
\bibitem{ch} In fact it has been shown that the large $r$ behavior of the stress-energy
for the massless spin $1/2$ field in the Extreme Reissner-Nordstr\"{o}m spacetime goes
like $M/r^5$:  E.D. Carlson and W.H. Hirsch, private communication.
\bibitem{chrisfull} S. M. Christensen and S. A. Fulling, {\it
Phys. Rev.} D{\bf 15}, 2088 (1977).
\bibitem{emparan} R. Emparan, private communication.
\bibitem{ahl}
P. R. Anderson, W. A. Hiscock and D. J. Loranz, {\it Phys. Rev. Lett.}
{\bf 74}, 4365 (1995) 
\bibitem{spin12} E.D. Carlson, W.H. Hirsch, B. Obermayer, P.R. Anderson and P. B. Groves, Phys. Rev. Lett.
{\bf 91}, 051301 (2003).

\end{references}
\end{document}